\documentclass[12pt,a4paper]{article}

\usepackage{amsmath}
\usepackage{amsfonts}
\usepackage{amssymb}

\usepackage{amssymb}
\usepackage{amsmath}
\usepackage{graphicx}
\usepackage[margin=5pt, font=small,labelfont=bf,justification=raggedright]{caption}

\usepackage{url}
\usepackage[bookmarks, pagebackref=false]{hyperref}
\usepackage{color}
	\definecolor{rossoCP3}{cmyk}{0,.88,.77,.40}
		\definecolor{graa}{rgb}{0.8,0.8,0.8}
		\definecolor{blaa}{rgb}{0.2,0.2,0.6}
		\definecolor{gron}{RGB}{0,150,0}
		\hypersetup{
			colorlinks, 
			bookmarksopen, 
			bookmarksnumbered,
			citecolor=blaa, 		
			linkcolor=black, 
			urlcolor=blaa,			
			}

\newcommand{\ea}[1]{
\begin{align}
#1
\end{align}
}

\newcommand{\lf}{\left}				\newcommand{\rg}{\right}

		\def\b{\beta}				
						\def\q{\theta}
						
			\def\x{\xi}			\def\p{\pi}			\def\r{\rho}	
		\def\t{\tau}						\def\y{\psi}
			\def\w{\omega}		\def\vf{\varphi}		

\def\G{\Gamma}		\def\D{\Delta}

		\newcommand{\calL}{{\mathcal L}}

\newcommand{\der}{{\partial}}										

\def\be{\begin{equation}}
\def\ee{\end{equation}}
\def\ba{\begin{eqnarray}}
\def\ea{\end{eqnarray}}
\def\bea{\begin{eqnarray}}
\def\eea{\end{eqnarray}}

\usepackage{color}

\title{ Probing Fuzzballs \\ with \\ Particles, Waves and Strings }

\author{M. Bianchi,  D. Consoli, J.F. Morales \\
Dipartimento di Fisica and I.N.F.N - Sezione di Roma 2\\
Universit\`a di Roma Tor Vergata, \\
Via della Ricerca Scientifica, I-00133 Roma, Italy\\
massimo.bianchi@roma2.infn.it, dario.consoli@roma2.infn.it, morales@roma2.infn.it
}

\begin{document}

\numberwithin{equation}{section}


\thispagestyle{empty}

\begin{titlepage}
\hfill MCTP-17-16
\vspace{1cm}

\begin{center}

{\bf {\Large  Probing Fuzzballs }
\vskip 0.3cm
{\Large with}
\vskip 0.3cm
{\Large   Particles, Waves and Strings}}\\

\vspace{1.5cm}

{\large Massimo Bianchi${}^{a,b}$, Dario Consoli${}^{a,b}$  and Jose Francisco Morales${}^b$}\\

\vspace{1cm}

{\it ${}^a$ Dipartimento di Fisica Universita di Roma ``Tor Vergata" }\\
 {\it ${}^b$I.N.F.N - Sezione di Roma ``Tor Vergata"}\\
{\it  Via della Ricerca Scientifica, I-00133 Roma, ITALY}\\

\vspace{14pt}

\end{center}

 \begin{abstract}
 
    We probe D1D5  micro-state geometries with massless particles, waves and strings. To this end, we study geodetic motion, Klein-Gordon equation and string scattering in the resulting gravitational background. Due to the reduced rotational symmetry, even in the simple case of a circular fuzzball, the system cannot be integrated elementarily.  Yet, for motion in the plane of the string profile or in the orthogonal plane to it, one can compute the deflection angle or the phase shift and identify  the critical impact parameter, at which even a massless probe is captured by the fuzzball if its internal momentum is properly tuned.  We find agreement among the three approaches, thus giving further support to the fuzzball proposal at the dynamical level.  
 \end{abstract}

\end{titlepage}

\setcounter{page}{1} \renewcommand{\thefootnote}{\arabic{footnote}}
\setcounter{footnote}{0}
\newpage

\tableofcontents

\section{Introduction}

The information loss paradox for  black holes can be satisfactorily addressed in String Theory relying on the Fuzzball proposal\cite{Lunin:2001jy}\nocite{Lunin:2002iz, Mathur:2003hj,Lunin:2004uu,Mathur:2005zp, Skenderis:2008qn}-\cite{Mathur:2008nj}.  In this framework, black hole micro-states are represented by smooth horizon-less geometries with the same asymptotic behaviour at large distance -- and thus the same `charges' -- as the putative black hole. Enormous success has been achieved in counting the micro-states for extremal 3- and 4-charged black hole states in  five and four dimensions  respectively \cite{Strominger:1996sh}\nocite{Breckenridge:1996sn,Maldacena:1996ky, Maldacena:1997de}-\cite{Maldacena:1998bw}. Identifying the corresponding geometries in the supergravity regime  has proved to be much harder   
\cite{Giusto:2004id}\nocite{Giusto:2004ip,Bena:2005va, Berglund:2005vb, Saxena:2005uk,Bena:2006kb,Bena:2007kg,Bena:2007qc,Giusto:2009qq,Giusto:2011fy,Lunin:2012gp,Giusto:2013bda,Gibbons:2013tqa,Bena:2015bea,Lunin:2015hma, Bena:2016agb, Bena:2016ypk,Pieri:2016cqz, Bianchi:2016bgx, Pieri:2016pdt}-\cite{Bianchi:2017bxl}. The micro-state geometries found so far cannot fully account for  the entropy  of `large' 3-charge black holes in five dimensions, let alone  the case of 4-charged black holes in four dimensions.      

More progress can be achieved by considering 2-charge systems in string theory. Four and five-dimensional `small' black-holes with two charges can be constructed in string theory in terms of closed oriented fundamental strings carrying Kaluza-Klein momentum  \cite{David:2006yn} (see  \cite{Elvang:2005sa}\nocite{Moura:2006pz, Sahoo:2006pm, Prester:2008iu,Bianchi:2010es}-\cite{Bianchi:2010dy} for studies of the  dynamical properties of these  systems). These black holes are not solution of Einstein gravity but of its higher derivative extension, so their horizons are of stringy size.  The stringy micro-states can be counted exactly and put in correspondence with regular geometries associated to the U-duality equivalent 2-charge system realised in terms of a bound-state of D1- and D5-branes. In this framework one can find smooth horizonless geometries of Einstein gravity in  six dimensions  characterised by a  string profile function that incarnate the fuzzball paradigm in a  fully successful way \cite{Mathur:2005zp, Skenderis:2008qn, Mathur:2008nj}.  This system will be the main subject of our investigations here. 

In this paper we explore the  fuzzball geometries by scattering massless closed string states from the D1/D5 brane system. We will follow closely the approach pioneered by Amati, Ciafaloni and Veneziano for high-energy string scattering in the Regge regime \cite{Amati:1987wq}\nocite{Amati:1987uf, Amati:1988tn, Amati:1988tn}-\cite{Amati:1990xe} that has later on been adapted to the study of closed-string scattering off D-branes \cite{Hashimoto:1996kf}\nocite{Hashimoto:1996bf, Garousi:1996ad,DAppollonio:2010krb,Bianchi:2011se, DAppollonio:2013mgj, DAppollonio:2013okd}-\cite{DAppollonio:2015oag}. We  present three different descriptions of the scattering process. First we consider the geodetic motion of a massless particle in the fuzzball gravitational background. Second, we  consider the `classical' wave scattering in the  fuzzball geometry. Finally, we study the scattering  of massless closed string states off the 2-charge D-brane bound-state. 

In the regime where the impact parameter $b$ is large compared to the typical scale $L_p$ of the geometry (`Schwarzschild radius'),
the quasi-elastic scattering process is characterised by a small deflection angle $\theta$ and it is dominated by soft processes in which a very large number of nearly on-shell gravitons are exchanged between the high-energy string state and the brane. The scattering is described by `ladder' diagrams  whose elementary block is the tree-level (disk) string-brane scattering amplitude {\it i.e.} the closed string two-point amplitude on the disk \cite{Hashimoto:1996kf}\nocite{Hashimoto:1996bf, Garousi:1996ad, DAppollonio:2010krb, Bianchi:2011se, DAppollonio:2013mgj, DAppollonio:2013okd}-\cite{DAppollonio:2015oag}. The deflection angle can be extracted  from the phase shift $\delta(E,b)$ describing the transition from the incoming to the outgoing asymptotic states\footnote{A general scattering process with incoming state $|i\rangle$  and outgoing state $\langle f|$ is  described  by the amplitude 
$
 {{T}}_{fi} ={ {\cal A}_{fi} \over \prod_{l=1}^{n} \sqrt{2E_l} }   
$ 
 with $E_l$ denoting the energies of both incoming and outgoing states and $ {\rm i} {{T}}_{fi}=\langle f|  (S-1) |i\rangle$  the  $T$-matrix element. In the case of elastic scattering of a single particle from a target one has a single incoming and outgoing state  with the same energy $E$, so ${{T}}_{fi} ={ {\cal A}_{fi} \over  2E  }  $. }
\be
e^{2 {\rm i} \, \delta (E,b)} 
=1+ { {\rm i} \, \widehat{\cal A}(E,b) \over  2E }    
\ee
with  $\widehat{\cal A}(E,b)$  the Fourier transform of the  string theory scattering amplitude.
 Alternatively, the phase shift $\delta (E,b)$ can be computed by comparing the solution of the field equation in the gravitational background at plus and minus infinite time. 
 
For closed-string scattering off a stack of Dp-branes, the relevant amplitude is a disk with two closed-string vertices in the bulk and no open-string insertions on the boundary  \cite{Hashimoto:1996bf}\nocite{Hashimoto:1996kf, Garousi:1996ad, DAppollonio:2010krb, Bianchi:2011se, DAppollonio:2013mgj, DAppollonio:2013okd}-\cite{DAppollonio:2015oag}. For scattering off D-brane bound-states, twisted open string insertions  on the boundary are required  to bind D1- and D5-branes \cite{Giusto:2009qq, Giusto:2011fy, Bianchi:2016bgx}. To be specific we will consider high-energy graviton scattering in the presence of two open-string fermionic moduli living at the D1-D5 intersection. The string  result will be shown to be in perfect agreement  with both geodetic deflection and classical wave scattering in the fuzzball geometry. We interpret this as a first step towards giving support to the fuzzball proposal at the dynamical level.  

The plan of the paper is as follows. In Section \ref{fuzzballgeo} we review the fuzzball geometry for a circular string profile.  In Section \ref{geodesics},  after reviewing the geodetic motion of massless probes of Dp-branes,  we study the motion in the fuzzball background. Contrary to the case of a single stack of branes, the geometry even in the simple case of a circular fuzzball has reduced rotational symmetry, and generically one cannot integrate the geodetic equation elementarily. Yet for motion in the 2-plane of the circle or in the orthogonal 2-plane the geodetic equations can be solved and the deflection angle computed.  Remarkably, for special choices of the kinematics these geodesics exhibit a peculiar behaviour: they get asymptotically trapped into `circular' orbits. These correspond to `critical' impact parameters, at which the massless probe is captured by the fuzzball. The fuzzball effectively behaves as a black hole for the selected channel. This behaviour is consistent with the idea that a black hole can be depicted as an ensemble of a large number of fuzzballs with all possible orientations, so that the probe will always find a large number of trapping fuzzball bits\footnote{We thank Guillaume Bossard for suggesting us this picture.}.   
 In Section \ref{wavescatt} we  will study the `classical' scattering of scalar waves governed by the Klein-Gordon equation in the fuzzball geometry, determine the phase-shift and compute the deflection angle as the derivative of the former with respect to the impact parameter. We show that a wave carrying only one of the two independent angular momenta turns out to `localise' in one of the two 2-planes. Once again we compute the deflection angle and  identify the critical impact parameter in this approach. Finally 
in Section \ref{stringscatt} we compute the corresponding string amplitude on a disk with mixed boundary condition and find perfect agreement with the other two independent approaches. 

Section \ref{conclusions} contains our conclusions and an outlook. 

\section{The fuzzball geometry}
\label{fuzzballgeo}

In order to construct the D1/D5 fuzzball, we start with the D1/D5 configuration 
\be
\begin{array}{ccccccccccccc}
 &    & 0 & 1 & 2 & 3 & 4 & 5 & 6 & 7 & 8 & 9  \\
 {\rm D1} &    & - & \cdot & \cdot & \cdot & \cdot  & - & \cdot & \cdot & \cdot & \cdot  \\
  {\rm D5} &    & - & \cdot & \cdot & \cdot & \cdot  & - & - &- & - & -  \\
\end{array}
\ee
where a line, respectively a dot, denote a Neumann (N), respectively a Dirichlet (D), direction. 
We will use the coordinates $(t,z)$ for the two NN directions $(X^0,X^5)$, $\vec{x}=(X^1,\dots X^4)$ for the ND directions $\vec y=(X^6,\dots X^9)$ for the remaining DD directions. We compactify $z$ and $\vec y$.

 The general  supersymmetric solution sharing with the D1/D5 system above  the  asymptotic geometry  is characterised by a metric of the form
\begin{equation}
ds^2_{\rm fuzz}=-(H_1 H_5)^{-1/2}[(dt+\vec{A}{\cdot}d\vec{x})^2-(dz+\vec{B}{\cdot}d\vec{x})^2]+(H_1 H_5)^{1/2}d\vec{x}{\cdot}d\vec{x}+\lf(\frac{H_1}{H_5}\rg)^{1/2} dy{\cdot} dy
\end{equation}
The 1-forms $A_1=\vec{A}{\cdot}d\vec{x}$ and $B_1=\vec{B}{\cdot}d\vec{x}$ are related by Hodge duality 
\begin{equation}
dA_1 =  *_4 dB_1
\end{equation}
and $H_i$ are harmonic functions of the $\vec{x}$ variables. In particular, regular solutions  can be constructed in this way by taking
\begin{align}
H_5(\vec{x})=1+\frac{L^2_5}{\lambda}\int_0^\lambda dv\, \frac{1}{|\vec{x}-\vec{F}(v)|^2} \\
H_1(\vec{x})=1+\frac{L^2_5}{\lambda}\int_0^\lambda dv\, \frac{|\dot{\vec{F}}(v)|^2}{|\vec{x}-\vec{F}(v)|^2}
\end{align}
where $\vec{F}(v)$ is the so-called profile function with values in $\mathbb{R}^4$. The 1-form $A_1$ has components
\begin{equation}
\vec{A}= {L_5^2 \over \lambda} \int_0^\lambda  dv\, \frac{\dot{\vec{F}}(v)}{|\vec{x}-\vec{F}(v)|^2}
\end{equation}
We will focus on the circular profile in the (1,2) plane:
\begin{equation}
\vec{F}(v)=a\lf(\cos \frac{2\p v}{\lambda},\sin \frac{2\p v}{\lambda},0,0\rg)
\end{equation}
with $\lambda =2 \p a L_5/L_1$. In order to compute the integrals, it is convenient to change coordinates and set
\be 
x_1+ix_2= \sqrt{  \rho^2+a^2 } \sin \vartheta  e^{i \vf} \quad , 
\quad x_3+ix_4= \rho \cos \vartheta e^{i \y}
\ee
In this coordinates the  metric  of a circular D1/D5 fuzzball reads
\begin{align}
\label{D1D5circfuzzmetric}
ds^2_{\rm fuzz}&{=}H^{-1}\lf[ -(dt+\w_\vf d{\vf})^2+(dz+ \w_\y d{\y})^2\rg] \\
&{+}H\lf[(\rho^2{+}a^2 \cos^2 \vartheta) \lf(\frac{d\rho^2}{\rho^2{+}a^2}{+}d\vartheta^2\rg){+}\rho^2 \cos^2\vartheta\, d {\y}^2{+}(\rho^2{+}a^2)\sin^2 \vartheta d {\vf}^2\rg] \nonumber
\end{align} 
where $H=\sqrt{H_1 H_5}$, with
\bea
H_i &=& 1+\frac{L^2_i}{\rho^2+a^2 \cos^2 \vartheta} \qquad , \nonumber \\
  \w_\vf &=&   \frac{a  L_1 L_5 \sin^2 \vartheta }{\rho^2+a^2 \cos^2 \vartheta} 
\qquad , \qquad
\w_\y = \frac{a  L_1 L_5  \cos^2 \vartheta}{\rho^2+a^2 \cos^2 \vartheta}  
\eea
 It has been shown that, despite the apparent singularities of $H_i$ along the string profile, the metrics defined in this way are regular everywhere. 

\section{Geodetic motion}
\label{geodesics}

In this section we study the geodetic motion of massless neutral particles in the fuzzball geometry. We start by reviewing the geodetic motion in the background of Dp-branes, paying particular attention to the D5-brane case for its relevance to the more interesting  fuzzball case, analysed afterwards. We will also identify the critical impact parameters at which even massless probes get trapped in the gravitational background of the fuzzball.

\subsection{Geodesics in the Dp-brane case}
\label{sect:Dp_brane}

We consider first the scattering from a stack of Dp-branes. The metric in the scattering plane transverse to the Dp-branes can be parametrised by the distance $r$ from the stack and an angle $\theta$. The metric in this plane takes the form
\begin{equation}
ds^2=-  H^{-{1\over 2}}(r) dt^2+H^{{1\over 2}}(r)(dr^2+r^2 d\q^2)
\end{equation}
The system, governed by $\calL= \frac{1}{2} \frac{ds^2}{d\t^2}$, admits two conserved quantities, energy and angular momentum, associated to  invariance under the shifts of the coordinates $t$ and $\q$:
\begin{equation}
-E=\frac{\der \calL}{\der \dot{t}}=-H^{-{1\over 2}}  \dot{t} \qquad , \qquad
J= \frac{\der \calL}{\der \dot{\q}} =H^{{1\over 2}} r^2 \dot{\q}
\end{equation}
where dots indicate derivatives wrt the affine parameter $\tau$.  In terms of these quantities, 
the geodesic equation for a massless particle, $ds^2=0$, becomes
\begin{equation}
 \dot{r}^2+\frac{J^2}{H r^2}-E^2\, H=0
\end{equation}
Writing $\dot{r} =\dot{\q} \frac{d r}{d \q}  = \frac{J}{\b r^2} \frac{d r}{d \q} $ one finds 
\begin{equation}
{d\q}=-  \frac{dr}{\sqrt{f(r)}}  \label{dthetar}
\end{equation}
where
\begin{equation}
f(r) = {H(r)  r^4 \over  \,b^2} -    r^2          \label{fdef}
\end{equation}
and $b=J/E$ is the impact parameter. We assume that the incoming particle arrives from the direction $\theta = 0$, thus the sign of the square root is minus because when $r$ decreases $\q$ increases. Eq. (\ref{dthetar}) can be integrated from $r=\infty$ up to $r_*$, the largest zero of $f(r)$. If $r_*$ is a simple zero of $f(r)$  {\it i.e.} the integral is finite. The particle reaches $r_*$, where the radial velocity vanishes $ \dot{r}(r_*)=0$, and bounces back to infinity following Eq. (\ref{dthetar}) with the opposite sign. We will refer to $r_*$ as the {\it turning point}. If $r_*$ is a higher-order zero of $f(r)$ 
the integral diverge and the particle gets trapped, looping around a {\it limiting cycle}. We will discuss this possibility later on.
 
Around the turning point the trajectory is symmetric and the total deflection angle is given by
\begin{equation}
\D \q=2 \q(r_*)-\p  \label{deflection}
\end{equation}
where
\bea
\q_*= \q(r_*) =    -\int_{+\infty}^{r_*}  dr  f(r)^{-{1\over 2} }  = r_*  \int_{0}^{1}  {d\xi \over \xi^2}  f\left( \frac{r_*}{\x} \right)^{-{1\over 2} }
\label{eq:theta_star_Dp}
\eea
where in the second line we used the variable $\xi=r_*/r$. 
 We are interested in the large $b$  limit. In this limit, assuming that $ H(r) \approx 1 + {\cal O}(1/r^n)$, one  finds
 \be
 f(r) ={r^4\over b^2}-r^2   +   {f_n\over   b^2\, r^{n-4}}  +\ldots   
 \ee
so that the turning radius is given by
\be
r_*=b \left( 1-   {f_n   \over 2 \, b^n} +\ldots   \right) 
\ee
and the turning angle reads
\begin{equation}
\q_*=\int_0^1 \frac{d\x}{\sqrt{1-\x^2}} \lf[1+\frac{1}{2}  {f_n\over b^n}  \frac{1-\x^{n}}{1-\x^2}+\dots\rg] ={\pi\over 2}+
{ \sqrt{\pi}  \,  \Gamma\left({n+1\over 2}\right) \over    2 \, \Gamma\left({n\over 2}\right) }\,{f_n\over b^n} +\ldots
\end{equation}
leading to
\be
\Delta \theta={ \sqrt{\pi}  \,   \Gamma\left({n+1\over 2}\right) \over    \Gamma\left({n\over 2}\right) }{f_n\over b^n}   \label{defang}
\ee
 In the specific case of a Dp-brane $n=7-p$ and 
\begin{equation}
 H(r) =1+\lf(\frac{L_p}{r}\rg)^{7-p}
\end{equation}
with $L_p^{7-p} = g_s N_p (2\pi\alpha')^{(7-p)/2}/\Omega_{8-p}$.  The deflection angle becomes \cite{DAppollonio:2010krb, DAppollonio:2013mgj}
\begin{equation}
\boxed{\D \q_{\rm Dp} = 
\sqrt{\p}\,\frac{\G(\frac{8-p}{2})}{\G(\frac{7-p}{2})}\lf(\frac{L_p}{b}\rg)^{7-p}+\dots}    \label{resdp}
\end{equation}
where dots denote subleading terms in ${L_p}/{b}$.

\subsection{Geodesics in the D5-brane case}

In preparation for the D1/D5 fuzzball, we will now consider in some more detail the  case of scattering from a stack of D5-branes. 
An  exact  formula for the scattering angle $\D \q$ can be obtained. In this case 
\be
H(r)=1+{L_5^2\over r^2}  
\ee
 and the turning point equation (\ref{turning}) is solved by  
\begin{equation}
r_*= \sqrt{b^2-L_5^2}
\end{equation}
 The turning point $r_*$ exists for $b\geq b_{\rm crit} =L_5$.  The deflection angle is given by  (\ref{eq:theta_star_Dp}-\ref{deflection})   leading to the exact formula
  \begin{equation}
\boxed{ \Delta \q_{\rm D5}=-\pi+{2\, b \over r_*} \,     \int_0^1 { d\xi \over  
 \sqrt{   1  -  \xi^2      }}
= \pi \,  \left[  {  b\over \sqrt{b^2-L_5^2} }  -1   \right] }\label{geod5}
\end{equation}
We notice that for $b$ near $b_{\rm crit}=L_5$, $\Delta \q\to \infty$, {\it i.e.} the particle trajectory describes an in-falling spiral around the brane stack.
    
%
%

\subsection{Geodesics in the D1/D5 fuzzball}

The metric (\ref{D1D5circfuzzmetric}) of a circular fuzzball has no explicit dependence on $t$, $z$, $\vf$ and $\y$. The system, governed by ${\cal L}={1\over 2} {ds^2\over d\tau^2}$, admits then four commuting isometries and as many invariants for the geodetic motion, {\it viz.} 
\begin{align}
-E&=\frac{\der \calL}{\der \dot{t}}=-\frac{1}{H}\lf(\dot{t}+\w_\vf \dot{\vf}\rg) \nonumber\\
P&=\frac{\der \calL}{\der \dot{z}}=\frac{1}{H} \lf(\dot{z}+ \w_\y \dot{\y}\rg) \nonumber \\
J_\vf&=\frac{\der \calL}{\der \dot{\vf}}= -\frac{\w_\vf}{H} (\dot{t}+\w_\vf \dot{\vf})+H (\rho^2+a^2)\sin^2 \vartheta \dot{\vf} \nonumber\\
J_\y&= \frac{\der \calL}{\der \dot{\y}}=\frac{\w_\y}{H} (\dot{z} + \w_\y \dot{\y})+H \rho^2 \cos^2 \vartheta \dot{\y}
\end{align}
that allow to determine the corresponding `velocities'
\begin{align}
\dot{\vf}&= \frac{J_\vf+\w_\vf E}{H(\rho^2+a^2)\sin^2 \vartheta}    \qquad~~~~~~~~~~~~
\dot{\y}= \frac{J_\y- \w_\y P}{H \rho^2 \cos^2 \vartheta} \nonumber\\
\dot{t} &= E\, H-\w_\varphi\, \dot{\varphi}  \qquad ~~~~~~~~~~~~~~~~~~~\dot{z}=P\, H-\w_\y \dot{\y}  \label{dots}
\end{align}
The   lightlike geodesic equation $ds^2=0$ becomes:
\begin{equation}
\label{fuzzgeod}
H (P^2-E^2)+H (\rho^2{+}a^2 \cos^2 \vartheta) \lf(\frac{\dot{\rho}^2}{\rho^2{+}a^2}{+}\dot{\vartheta}^2\rg){+}
 \frac{(J_\y- \w_\y P)^2}{H \rho^2 \cos^2 \vartheta}{+} \frac{(J_\vf+\w_\vf E)^2}{H (\rho^2+a^2)\sin^2 \vartheta}=0
\end{equation}
Despite the reduced isometry, $U(1)_\psi \times U(1)_\varphi \subset SO(4)$, the system is separable\footnote{We thank Yuri Chervonyi and Oleg Lunin for drawing these results to our attention.} \cite{Cvetic:1996xz,Cvetic:1997uw,Lunin:2001dt,Chervonyi:2013eja}.  In order to expose this property, one introduces the conjugate momenta 
 \bea
 \label{PrhoPtheta}
 P_{\rho} &=&\frac{\der \calL}{\der \dot{\rho}}={ H (\rho^2{+}a^2 \cos^2 \vartheta) \dot{\rho} \over {\rho^2{+}a^2}}  \nonumber\\
P_{\vartheta} &=& \frac{\der \calL}{\der \dot{\vartheta}}={ H (\rho^2{+}a^2 \cos^2 \vartheta) \dot{\vartheta}}
 \eea
 Expressing $ \dot{\rho}$ and $\dot{\vartheta}$ in terms of $P_{\rho}$ and $P_{\vartheta}$, the geodesic equation  (\ref{fuzzgeod}) can be written in the form
\bea
\label{fuzzgeodsep}
&&H^2 (\rho^2{+}a^2 \cos^2 \vartheta) (P^2-E^2) + ({\rho^2{+}a^2})P_{\rho}^2 + P_{\vartheta}^2 \\
 &&\quad + \frac{(\rho^2{+}a^2 \cos^2 \vartheta)(J_\y- \w_\y P)^2 }{\rho^2 \cos^2 \vartheta}{+} \frac{(\rho^2{+}a^2 - a^2\sin^2 \vartheta)(J_\vf+\w_\vf E)^2}{(\rho^2+a^2)\sin^2 \vartheta}=0 \nonumber
\eea
that can be separated into two equations \bea
\label{fuzzgeodsepK}
 \lambda&=&P_{\vartheta}^2 + {J_\y^2 \over \cos^2\vartheta} + {J_\vf^2 \over \sin^2 \vartheta} - a^2 \cos^2 \vartheta
(E^2-P^2)   \\
 {-}\lambda&=&
 ({\rho^2{+}a^2})P_{\rho}^2+{(P L_1 L_5{-}a J_\y )^2  \over\rho^2}-{(E L_1 L_5 {-} a J_\vf)^2 \over\rho^2 + a^2}- 
 ({\rho}^2{+}L_1^2{+}L_5^2)(E^2{-}P^2)  \nonumber
\eea
where $\lambda$ is a separation constant, that can be conveniently expressed as $\lambda=K^2-a^2 (E^2-P^2)$.  The geodesics in the $(\vartheta,\rho)$ plane is described by the equation
\be
\label{dthesudrho}
{d\vartheta \over d\rho} = {\dot\vartheta \over \dot\rho}  = {P_\vartheta (\vartheta; \mathfrak I)  \over (\rho^2+a^2) P_\rho(\rho; \mathfrak I)}
\ee
with  $\mathfrak I=\{K, E, P, J_\y, J_\vf\}$ collectively denoting the constants of motion and
 \bea
 \label{Ptheta}
  P_\vartheta (\vartheta; \mathfrak I)^2  &=&  K^2 - a^2 \sin^2 \vartheta
(E^2-P^2) - {J_\y^2 \over \cos^2\vartheta} - {J_\vf^2 \over \sin^2 \vartheta} \\
P_\rho(\rho; \mathfrak I)^2  &=&   
\left( 1{+} {L_1^2{+} L_5^2\over\rho^2 {+}  a^2}\right) (E^2{-}P^2){-}
{(P L_1 L_5{-} a J_\y )^2  \over\rho^2({\rho^2{+}a^2})}{+}{(E L_1 L_5 {-}  a J_\vf)^2 \over(\rho^2 {+} a^2)^2} 
{-}{K^2\over\rho^2 {+}  a^2}   \nonumber
   \eea
The general solution of (\ref{dthesudrho}) requires the inversion of (incomplete) elliptic integrals that is beyond the scope of the present investigation. 

Since the system cannot be integrated elementarily, 
for simplicity and illustrative purposes,
we focus on the case of geodesics at constant  $\vartheta= \vartheta_0$, {\it i.e.} $\dot{\vartheta}=0=P_\vartheta$. Equation (\ref{Ptheta}) yields
\begin{equation}
\label{thetaconst}
K^2 = a^2 \sin^2 \vartheta_0
(E^2-P^2) + {J_\y^2 \over \cos^2\vartheta_0} + {J_\vf^2 \over \sin^2 \vartheta_0}
\end{equation}
This fixes $K$ in terms of the other conserved quantities and the incident angle $\vartheta_0$.

For simplicity, in the following, we will restrict our attention on these two choices:
\bea
\vartheta_0  &=& 0  \qquad      J_\varphi= \w_\varphi =0   \qquad K^2=J_\y^2  \nonumber\\
\vartheta_0  &=& {\pi\over 2}  \qquad      J_\psi=\w_\psi  =0  \qquad K^2=J_\vf^2 + a^2(E^2-P^2) 
\eea

\subsubsection{ Geodesics in the plane $\vartheta=0$}
 
 In this plane  the   light-like geodesic equation $ds^2=0$ becomes:
\begin{equation}
\dot{\rho}^2  = E^2-P^2-  \frac{(J_\y- \w_\y P)^2}{H^2 \rho^2  } 
\end{equation}
To find $\psi$ as a function of $\rho$ we write  
\be
\dot{\rho}=\dot{\psi} \frac{d \r}{d \psi} = \frac{J_\y- \w_\y P}{H \rho^2  } \frac{d \r}{d \psi} 
\ee
 leading to
\be
\frac{d\rho}{d\y} =-f(\rho)^{{1\over 2}}=
  -\left[  {   \left( \rho^2+a^2+L_1^2 \right)   \left( \rho^2+a^2+L_5^2 \right)    \over  b^2 \lf( \rho^2+a^2 -  {a\, v \,L_1\, L_5 \over b_\y  }   \rg)^2 }  \, \rho^4  -\rho^2 \right]^{{1\over 2}}       \label{dthetar2} 
\end{equation} 
 with
 \be
       v={P\over E}\quad ,  \quad    b_\y={J_\y \over E} \quad ,  \quad  b={b_\psi \over \sqrt{1-v^2} }
 \ee
 At large $\rho$ one finds
  \be
 f(\rho)=   {\rho^4 \over b^2}-\rho^2+  {f_2\, \rho^2 \over b^2}+\ldots        \label{rholarge}
 \ee
 with
 \be
 f_2= \left( L_1^2+L_5^2  +2 
 \frac{a \, v L_1 L_5}{ b_\y} \right) 
 \ee
  Following the same steps as before one finds a turning point located at 
 \be
 \rho_*=b \left(1-{f_2\over 2\, b^2} \right)
 \ee
 and the deflection angle
\begin{equation}
\boxed{ \D \y_{\rm fuzz} = \frac{\p (1-v^2)}{2 b_\psi^2 }
 \left( L_1^2+L_5^2  +2 
 \frac{a \, v L_1 L_5}{ b_\y} \right) +\dots }
\end{equation}

\subsubsection{  Geodesics in the plane $\vartheta=\pi/2$}

In this plane  the   lightlike geodesic equation $ds^2=0$ becomes:
\begin{equation}
H (P^2-E^2)+{H \, \rho^2\,\dot{\rho}^2 \over \rho^2+a^2}  + \frac{(J_\vf+\w_\vf E)^2}{H (\rho^2+a^2) } =0
\end{equation}
To find $\varphi$ as a function of $\rho$ we write
\begin{equation}
\dot{\rho}=\dot{\varphi} \frac{d \r}{d \varphi} =   \frac{J_\varphi  +  \omega_\varphi E     }{H  (\rho^2+a^2)  }  \frac{d \r}{d \varphi}
\end{equation}
leading to 
\begin{equation}
\frac{d\rho}{d\vf}=-f(\rho)^{1 \over 2}= -  \left[   {    \left( 1+{L_1^2 \over \rho^2} \right)  \left( 1+{L_5^2 \over \rho^2} \right) \left(1+{a^2\over \rho^2}\right)^3   
 \over  b^2  \left( 1 +{ a  L_1 L_5  \over
     b_\vf \, \rho^2 }   \right)^2}   \, \rho^4  -\rho^2 \, \left(1+{a^2\over \rho^2} \right)^2    \right]^{{1\over 2}}  
\label{dthetar0}
\end{equation}   
with
\begin{equation}
  v={P\over E}    \quad  , \quad      b_\vf={J_\vf \over E}  \qquad ,  \qquad  b={b_\vf \over \sqrt{1-v^2}} 
\end{equation}
At large $\rho$ we find
\begin{equation}
    f(\rho)= {\rho^4\over b^2}-\rho^2+ {f_2\, \rho^2 \over b^2}-2 \, a^2+\ldots   
    \ee
    with
    \be
    f_2=  \left( L_1^2+L_5^2  -2 
 \frac{a \,  L_1 L_5 }{ b_\vf }+ 3\, a^2 \right)
\end{equation}
Now the turning point is located  at
\be
\rho_*=b\left(  1+{ 2a^2- f_2 \over 2 \, b^2} +\ldots \right)
\ee
  The turning angle becomes
\bea
\vf(\rho_*) &=&        \int_{0}^{1}  d\xi \,   \left( {\rho_*^2\over b^2} -{  \xi ^2}+{f_2\, \xi^2 \over b^2 } -{ 2 a^2 \, \xi^4 \over \rho_*^2} \right)^{-{1\over 2} }  \nonumber\\
&=&\int_0^1 \frac{d\x}{\sqrt{1-\x^2}} \lf[1+   { f_2-2a^2(1+\xi^2) \over 2\, b^2}  +\dots\rg] \nonumber\\
&=& {\pi\over 2}+{\pi\over 4 b^2} (f_2-3\, a^2)+\ldots
\eea
leading to
 \begin{equation}
\boxed{\D\vf_{\rm fuzz} = \frac{\p(1-v^2)}{2 b_\vf^2}
 \left( L_1^2+L^2_5  -2 
 \frac{a \,  L_1 L_5 }{ b_\vf }  \right) + \dots}
\end{equation}

 \subsubsection{Critical impact parameter  }
 
 So far we  have considered scattering at such a large impact parameter $b$ that the particle has enough kinetic energy to escape from the gravitational potential without hitting the target. In this limit, a turning point is found at $\rho_* \approx b>0$. Indeed at $\rho=\rho^*$, $\dot{\rho}=0=P_\rho$ and a finite deflection is observed. In general, a turning point exists if the largest real zero of the numerator of $P^2_\rho$ is a simple root. On the other hand, a multiple such zero, where $f(\rho^*) = 0 = f'(\rho^*)$, signals a limiting cycle along which even a massless particle gets asymptotically trapped.  
 
 In the following we will show that for small enough $b_{\varphi/\psi}$ a critical value of the internal momentum $P$ exists such that the null geodesic is trapped. This may sound somewhat surprising since the fuzzball geometry is regular, unlike Dp-branes or black holes, and we do not expect that massless particles or light get captured in a regular gravitational background. Indeed, as we will see, for generic values of the constants of motion a turning point always exists
for any choice of the impact parameter different from the critical value. 
 
We will restrict our attention on the two special cases $\vartheta=0,\pi/2$ and in order to further simplify our analysis we set $L_1=L_5=L$ and take
 \be
 P = {a J_\psi\over L^2 }
 \ee
in the following. For this choice $P_\rho^2$ becomes a quadratic polynomial in $\rho^2$ up to a nowhere vanishing function and the critical impact parameter $b_{\rm crit}$ can be explicitly determined.  
 
 \subsubsection*{Case $\vartheta=0$}
 
A turning point $\rho_*$ in the geodesics at $\vartheta=0$ exists if and only if the largest positive zero of the function $f(\rho)$ 
\be
 f(\rho)=   {    \left( \rho^2+a^2+L^2 \right)^2     \over  b^2 \lf( \rho^2+a^2 -  {a\, v \,\, L^2 \over b \sqrt{1-v^2}  }   \rg)^2 }  \, \rho^4  -\rho^2      \label{frho}
\ee
is simple.  We notice that  for generic values of $v$, the function $f(\rho)$ behaves as
 \be
 f(\rho) \approx 
\left\{
\begin{array}{ccc}
 { \rho^4 \over b^2}>0  & {\rm for}  & \rho\to \infty   \\
  -\rho^2<0  & {\rm for}  & \rho\to 0   \\
\end{array}
\right.   \label{finf}
 \ee
Consequently, $f(\rho)$ should have a zero $\rho_*$ on the positive real axis and therefore a turning point exists for generic values of $v$. 
On the other hand for our choice $P L^2=a J_\y$ {\it i.e.}
\be
 v= \left( 1+{L^4\over a^2 b^2} \right)^{-{1\over 2}} 
\ee
the function $f(\rho)$ is positive both for $\rho$ large or small so the existence of a zero cannot be taken for granted. The turning point equation reduces to
\be
      \rho_*^2 -b\, \rho_* +a^2+L^2   =0
\ee    
   which has no solutions if $b<b_{\rm crit}$ with
\be
\boxed{ b_{{\rm crit}}=2 \sqrt{L^2+a^2} }
\ee   
At $b=b_{{\rm crit}}$ the massless probe reaches asymptotically a `circular' orbit 
\begin{equation}
d\y= - \frac{2 \r_{{\rm crit}} \, d\r}{\r^2-\r^2_{{\rm crit}}}
\end{equation}
at a radius $ \r_{{\rm crit}}= b_{\rm crit}/2$,  and winds around it an infinite number of times, thus getting trapped for ever. 

Below $b_{{\rm crit}}$ the probe can reach $\r=0$, pass through and escape to infinity, possibly after looping few times. 
 
  \subsubsection*{Case $\vartheta=\p/2$}

  A turning point $\rho_*$ in the geodesics at $\vartheta={\pi \over 2}$ exists if and only if, the largest positive zero of the function $f(\rho)$ 
\be
    f(\r)=\frac{\left(\r^2+a^2 \right)^2}{\r^2}  \left[{  \left( \r^2+L^2 \right)^2 \left(\rho^2+a^2\right)   
 \over  b^2  \left( \rho^2 +{ a  L^2  \over
     b\sqrt{1-v^2}}   \right)^2}    -1 \,   \right]     \label{frho}
\ee
is simple. For our choice $v=0$, the turning point equation becomes
 \begin{equation}
 \r_*^4+(2L^2{+}a^2{-}b_\vf^2)\r_*^2+L^2(L^2+2a^2-2a b_\vf)=0
 \end{equation}
whose zeroes are given by
\begin{gather}
\r^2_\pm = \frac{b_\vf^2{-}2L^2{-}a^2}{2}\pm \frac{\sqrt{\D}}{2}\qquad {\rm with} \qquad 
\D=(b_\vf{-}a)^2 (b_\vf{+}a{+}2L)(b_\vf{-}a{+}2L)
\end{gather}
The turning point exists when $\D\geq 0$ and $\r^2_+ \geq 0$ thus the critical impact parameter can be obtained when $\D=0$ and $\r_+^2\geq 0$ or $\D\geq0$ and $\r^2_+=0$. The solutions are\footnote{$b_\vf<0$ means $J_\vf<0$, {\it i.e.} the probe is counter-rotating with respect to the fuzzball.}
 \be
\boxed{ 
b_{{\rm crit}}^-  =  -2 L-a  \qquad
b_{{\rm crit}}^+  =    
\left\{
\begin{array}{ccc}
{L^2\over 2a}+a   & {\rm for}  & L\leq 2 a  \\
 2 L-a  & {\rm for}  & L\geq 2 a  \\
 \end{array}
\right.}
 \ee
As in the $\vartheta=0$ case, when $b=b_{{\rm crit}}^\pm$ the massless probe gets trapped in an asymptotic `circular' orbit. In between these values of $b$ the particle can reach $\r=0$, that corresponds to the circular profile, and escape to infinity, possibly after looping few times.

\section{Classical scalar wave scattering}
\label{wavescatt}

In this section we consider the scattering of a classical wave in the fuzzball background. Again we start by considering the Dp-brane case with particular attention on the D5-brane case. We compute the scattering matrix, the phase shift and the deflection angles for the harmonic modes building the incoming plane wave.  To each harmonic mode we  can associated a classical geodesics. In particular, we will show that harmonic modes. We show that a wave carrying only one of the two independent angular momenta turns out to `localise' in one of the two 2-planes with a deflection angle that matches its particle analog.

\subsection{The scattering amplitude}

 The geodetic motion of a massless spin-less particle  on a gravitational background can be alternatively described by a massless scalar waves $\Phi({\bf  x},t ) $ satisfying the Klein-Gordon (KG) equation 
  \be
\square \,\Phi({\bf x},t ) ={1\over \sqrt{g}} \partial_\mu \left[ \sqrt{g} \, g^{\mu\nu} \, \partial_\nu \Phi({\bf x},t )  \right] =0   
\ee
Elastic scattering at energy $E$ from a point-like stationary target located at the origin 
is described by a scalar wave of the form\footnote{In a classical relativistic process such as the one under consideration, one should put $\omega = E/\hbar$. We choose units such that $\hbar=c=1$. }
 \be
\Phi({\bf x},t )=e^{-{\rm i}\, E \, t} \, \Phi({\bf x})
\ee
and the KG equation becomes
\be
  \left[     \nabla_{\rm flat}^2    +E^2  -\widehat{U}_{\rm eff} \right]  \Phi ({\bf x}) = 0    \label{box}
\ee
with  $\nabla_{\rm flat}^2$ the Laplacian computed with the flat space metric
  \be
 ds^2_{\mathbb R^d} =dr^2+r^2 (d\theta^2+\sin^2\theta\, d\Omega_{d-2})
 \ee
 and $\widehat{U}_{\rm eff} = \nabla^2 - \nabla_{\rm flat}^2 $   a differential operator encoding the details of the gravitational potential. For Dp-branes, $\widehat{U}_{\rm eff}$ depends only on $r$ and derivative thereof. For this reason, we will denote it by $\widehat{U}_{\rm eff} (r)$.
Using rotational symmetry, it is convenient to expand the scalar wave in spherical harmonics  
  \be
\Phi({\bf x})  =   \sum_{\ell=0}^\infty  {{R}}_\ell(r)  \, C^\alpha_\ell (\cos \theta) 
\ee
 with  
 \be \alpha={d-2\over 2}
 \ee
and $C^\alpha_\ell (x)$ the Gegenbauer polynomials (see Appendix \ref{sect:spherical_harmonics} for details)
satisfying
 \be
\nabla^2_{S^{2\alpha+1}}\,  C^{\alpha}_\ell (\cos \theta) =- \ell(\ell+2\alpha) \, C^{\alpha}_\ell (\cos \theta)
\ee 
The $\ell$-component ${{R}}_\ell(r)$ satisfies the radial wave equation
\bea
  \left[  {1\over r^{2\alpha+1} } \partial_r ( r^{2\alpha+1}\, \partial_r )    - {\ell(\ell+2\alpha)\over r^2}     +E^2 -\widehat{U}_{\rm eff}(r)   \right]  {{R}}_\ell (r) = 0  
  \eea
  We decompose the solution into a sum of a free incoming $\Phi_{\rm in}({\bf x})$  and a scattered wave $\Phi_{\rm out}({\bf x})$ 
   \be
\Phi({\bf x})  =\Phi_{\rm in}({\bf x}) +\Phi_{\rm out}({\bf x})  = \sum_{\ell=0}^\infty  \left[ {{R}}_{{\rm in} ,\ell}(r) +{{R}}_{{\rm out} ,\ell}(r)   \right]\, C^\alpha_\ell (\cos \theta) 
\ee
  with the incoming radial  harmonics being solutions of the D'Alambertian equation in flat space
  \bea
  \left[  {1\over r^{2\alpha+1} } \partial_r ( r^{2\alpha+1}\, \partial_r )    - {\ell(\ell+2\alpha)\over r^2}     +E^2    \right]  {{R}}_{ {\rm in},\ell} (r) = 0  
  \eea
 The Klein-Gordon equation (\ref{box}) can be conveniently written in the equivalent integral form
   \be
  {{R}}_\ell (r)=  {{R}}_{{\rm in},\ell}(r) +   \int_0^\infty  G_\ell^1 (r,r') \, \widehat{U}_{\rm eff} (r') \, {{R}}_\ell (r') \, r'^{\,2\alpha+1} dr'   \label{intform}
 \ee
  with  $G_\ell^a (r,r')$, $a=1,2$, denoting the flat-space (advanced/retarded) radial Green functions that satisfy 
  \be
   \left[  {1\over r^{2\alpha+1} } \partial_r ( r^{2\alpha+1}\, \partial_r )  - {\ell(\ell+2\alpha)\over r^2}     +E^2    \right]  G^a_\ell (r,r') = {\delta(r-r') \over r^{2\alpha+1}} 
\ee
 The general solution of the homogenous equation and the  Green function can be written in terms of Hanckel functions of the first and second kind
  \be
H^{1}_{n}(z)=   J_{n}(z) + {\rm i}\,  Y_{n}(z)   \quad , \quad H_\ell^{2}=( H_\ell^{1 })^*
\ee
where $J_n$ ($Y_n$) denote Bessel (Neumann) functions. 
 More precisely, the general solution of the homogenous equation can be written as
  \be
   {{R}}_{{\rm in},\ell}(r) = \sum_{a=1}^2 c_{\ell a} {H^a_{\ell+\alpha}(E\, r  )  \over r^\alpha}  \label{incoming}
  \ee
with $c_{\ell a}$ some coefficients. The Green function in flat space reads \cite{partialwave}
  \be
 G^a_\ell(r,r') =-{\rm i} {\pi \,   \over 2 } \, { J_{\ell+\alpha} (E \,r_{\rm min} ) \over r_{\rm min}^\alpha}     \,   {  H^a_{\ell+\alpha}(E\, r_{\rm max} )  \over r_{\rm max}^\alpha} 
 \label{green}
  \ee 
 with $r_{\rm min}$ and $r_{\rm max}$ the smallest and largest respectively of $r$ and $r'$. At infinity
    \be
  { H^{1}_{\ell+\alpha}( E\, r)   \over r^{\alpha} }
  \underset{ E r \to \infty }{\approx } \sqrt{2\over \pi E} { e^{ {\rm i} E r- {{\rm i}\pi \over 2} \left(\ell+\alpha  \right) } \over  r^{   \alpha+{1\over 2}   } } +\ldots
    \ee
so the asymptotic solution (\ref{incoming})  can be viewed in general as a superposition of an incoming and an outgoing wave. The  scattering process is described by an S-matrix computing the phase shift of the outgoing wave.  We choose the incoming wave to be a plane wave
 \be
  {{\Phi}}_{{\rm in}}({\bf x})  =  e^{{\rm i} E r \cos\theta}  = \sum_{\ell=0}^\infty  {{R}}_{{\rm in},\ell}(r)  \, C^\alpha_\ell (\cos \theta) 
\ee
with
\be  
  {{R}}_{{\rm in},\ell}(r)   = { 2^{\alpha } \, \Gamma(\alpha) \, {\rm i}^\ell (\ell+\alpha) \over E^\alpha}      {   J_{\ell+\alpha}(E\, r  )  \over r^\alpha}  \label{incoming}
    \ee
 The outgoing scattered  wave can be written as 
  \be
 \Phi_{{\rm out}}({\bf x}) =  { 2^{\alpha-1 } \, \Gamma(\alpha)  \over (E r)^\alpha}      \sum_{\ell=0}^\infty    {\rm i}^\ell (\ell+\alpha)  \,   ( e^{2\,{\rm i}  \delta_{\ell,E}}-1)   \,   H^1_{\ell+\alpha}(E\, r  )    \, C^\alpha_\ell (\cos \theta) 
 \label{phiout}
 \ee
 At infinity
 \be
  \Phi_{{\rm out}}(r) \approx  \,   f(\theta,E)\,   {e^{ {\rm i} E r - {{\rm i}\pi  \alpha \over 2}  } \over    r^{   \alpha+{1\over 2} }    } 
  \ee
  with
  \be
f(\theta,E)=     {2^{\alpha } \, \Gamma(\alpha) \,    \over  E^{\alpha+{1\over 2}} \sqrt{2 \pi }     }   
\sum_{\ell}^\infty      (\ell+\alpha) \,    e^{2\,{\rm i}  \delta_{\ell,E}}     \, C^\alpha_\ell (\cos \theta) 
\ee
 the scattering amplitude\footnote{ Here we used the identity
$ \sum_{\ell=0}^\infty      (\ell+\alpha) \,      \, C^\alpha_\ell (\cos \theta) =0$. We normalise $f$ in such a way that $d\sigma/d\Omega = |f|^2$. In $d=2\alpha+2$ spatial dimensions $\sigma  \sim L^{2\alpha +1}$ and $f \sim L^{\alpha + {1\over 2}}$. This is related to the (string) scattering amplitude ${\cal A}(s,t)$ by $f=E^{\alpha - {1\over 2}}{\cal A}$. }. 
In the high energy limit, the sum over $\ell$ can be viewed as an integral over $b=\ell/E$ after identifying $db=\Delta\ell/E$.  Moreover in this limit, the sum 
  is dominated by large $\ell$ modes and one can replace the spherical harmonics by its asymptotic form
\be
 C^\alpha_\ell (\cos \theta) \approx {2^{1-\alpha} \, \ell^{\alpha-1}\over (\sin\theta)^\alpha \, \Gamma(\alpha)}  \cos\left[  \theta\, (\ell+\alpha)-{  \pi\, \alpha\over 2}  \right]
\ee 
  One is left with
\be
 f(\theta,E)    \approx   { \sqrt{2 E } \,    \over  \sqrt{\pi} \theta^\alpha   } 
 \int_0^\infty db \,   b^\alpha  \,  e^{2\,{\rm i}  \delta(b,E)}     \,  \cos\left(  \theta\, b E {-}{  \pi\, \alpha\over 2}  \right) 
 \ee
 or equivalently\footnote{
One can rewrite the integral  over $b$ as  a $2\alpha+1$-dimensional integral using
\bea
&&\int d^{2\alpha+1} {\bf b} \, e^{{\rm i} \, {\bf q}\, {\bf b}+2 {\rm i} \delta(b) } \,  = \Omega_{2\alpha-1} \int db  \, b^{2\alpha}\, 
d\theta '  \,   \sin^{2\alpha-1} \theta' \, e^{{\rm i} \, q\, b\, \cos\theta'+2 {\rm i} \delta(b)  } 
\nonumber\\
&& =   \sqrt{\pi}   { \Omega_{2\alpha-1}  \, 2^{\alpha-{1\over 2} }  \Gamma\left(\alpha \right)  \over q^{\alpha -{1\over 2} }} \,
\int db \,   b^{\alpha+{1\over 2} }  \,  e^{ 2 {\rm i} \delta(b)  }  J_{\alpha}(q b)     
\approx   2\, \left({ 2 \pi \over q} \right)^{ \alpha  }          \, 
\int db \,   b^{\alpha  }  \, e^{ 2 {\rm i} \delta(b)  }\,   \cos \left( \theta \,b \, E  -{\pi \, \alpha \over 2}  \right)    
\nonumber \eea
 with  $\Omega_{n} =2 \pi^{n+1\over 2}/\Gamma\left( {n+1\over 2}\right)  $ the volume of the unit $n${-}sphere $S^n$.   }
 \be\label{wavescattamp}
 \boxed{
 f(\theta,E)    \approx   E^{\alpha+{1\over 2}}  
 \int  { d^{2\alpha+1} {\bf b}  \over (2\pi)^{\alpha+{1\over 2} }}\, e^{{\rm i} \, {\bf q}\, {\bf b}+2 {\rm i} \delta(b) }  
 }
\ee
 with $q=2 E \sin{\theta\over 2} \approx E \theta$.    The integral is dominated by a saddle point, let us say, at
\be
\boxed{
\theta=- {2\over E } {\partial \delta(E,b)  \over \partial b}   
}
 \label{thetadelta}
\ee
 This equation relates the deflection angle $\theta$ to the scattering amplitude $S=e^{ 2\,{\rm i}\delta(E,b)}$. 
 
 On the other hand, plugging (\ref{green}) into (\ref{intform}) one finds
 \be
  R_{{\rm out},\ell}(r)   = -{{\rm i}\pi\over 2} { H^1(E r)\over r^\alpha}  \int_0^\infty  J_{\ell+\alpha} (E \,r' )     \, \widehat{U}_{\rm eff} (r') \, {{R}}_\ell (r') \, (r')^{\alpha+1} dr' 
 \ee
 Comparing with (\ref{phiout})   one finds
  \be
  \boxed{
 e^{2\,{\rm i}  \delta_{\ell,E} } -1 =  - { {\rm i}  \pi E^\alpha \over 2^{\alpha } \, \Gamma(\alpha) {\rm i}^\ell (\ell+\alpha) } \,   \int_0^\infty  J_{\ell+\alpha} (E \,r' )     \, \widehat{U}_{\rm eff} (r') \, {{R}}_\ell (r') \, (r')^{\alpha+1} dr'   \label{phaseshift}
 }
 \ee

For instance, let us consider an effective potential
   \be
 \widehat{U}_{\rm eff} (r')=- {c_n \over r^n}
 \ee
    with $c_n$ small. In the limit where $c_n$ is small, the leading contribution to the phase shift  comes from replacing the radial function ${{R}}_{\ell}$ inside the integral in (\ref{phaseshift}) with its asymptotic value  ${{R}}_{ {\rm in}, \ell}$ so that
     \bea
 && e^{2\,{\rm i}  \delta(b,E) } -1  \approx  2\,{\rm i}  \delta(b,E)  \approx  {\rm i} \pi  \, c_n    \int_0^\infty  J_{b\, E+\alpha}^2 (E \,r' )  \, (r')^{1-n} dr'   \nonumber \\
  &&\approx     
  {\rm i} { \sqrt{\pi}  \, c_n \over 2}  {  E^{n-2} \Gamma\left(  { n-1\over 2} \right)   \Gamma\left(  b\, E+1\right)   \over 
 \Gamma\left(  { n\over 2} \right)   \Gamma\left(  b\, E+n\right)    } 
 \approx  {\rm i} {\sqrt{ \pi}  \, c_n \,    \over  2 \, E\,  b^{n-1}  }  
 {    \Gamma\left(  { n-1\over 2} \right)    \over 
 \Gamma\left(  { n\over 2} \right)      } \label{phaseshift2}
 \eea
    where in the last line we made use of Stirling formula for Euler's $\Gamma$-functions at large $b E$. 
    For the deflection angle one finds
    \be
    \boxed{
\Delta \theta=- {2\over E } {\partial \delta(E,b)  \over \partial b}  \approx    { \sqrt{ \pi}  \, c_n \,    \over    E^2\,  b^{n}  }  
 {    \Gamma\left(  { n+1\over 2} \right)    \over 
 \Gamma\left(  { n\over 2} \right)      }  
 }
 \label{thetadeltacn}
\ee
    
  \subsection{Dp-brane case}
 
  Let us consider  a Dp-brane gravitational background
 \be
ds^2=-H(r)^{-{1\over 2}}\, dt^2+H(r)^{1\over 2} \, (dr^2+r^2\, d\Omega_{8-p} )   \qquad , \qquad H(r) =1+ \left( {L_p\over r }\right)^{7-p}
\ee
The wave equation becomes
\be
   \left[    {1\over r^{3}} \partial_r (r^{3}\, \partial_r) + {1\over r^2} \nabla_{S^{8-p}}^2    +E^2 \, H +  { \partial_r  H^{6-p\over 4}  \over H^{6-p\over 4} } \partial_r  \right]  \Phi({\bf x})=0
\ee
 The last term is subleading in the limit of large $r$ and can be discarded, so one is left with
 \be
   \left[    {1\over r^{3}} \partial_r (r^{3}\, \partial_r) + {1\over r^2} \nabla_{S^{8-p}}^2    +E^2  + E^2 \left( {L_p\over r }\right)^{7-p}  \right]  \Phi({\bf x})=0 \label{eqondad5}
\ee
 The effective potential  is then
 \be
 \widehat{U}_{\rm eff} (r')=-E^2 \left( {L_p\over r }\right)^{7-p}
 \ee
 with $n=2\alpha=7-p$. Plugging this into (\ref{thetadeltacn})  one finds
 \begin{equation}
 \boxed{
\D \q_{\rm Dp} =  \sqrt{\p}\, \frac{\G(\frac{8-p}{2})}{\G(\frac{7-p}{2})}\lf(\frac{L_p}{b}\rg)^{7-p}+\dots   
}
 \label{resdp}
\end{equation}
 in agreement with (\ref{resdp}). 
 
  \subsection{D5-brane case}
 
In the case of a stack of D5-branes, the effective potential  is  
 \be
 \widehat{U}_{\rm eff} (r')=-{ E^2\, L^2 \over r^2}  
 \ee
 and equation (\ref{eqondad5}) has exactly the same form as the free equation with $\ell$ replaced by $\ell_L$ defined by
 \be
 -\ell_L(\ell_L+2)  =   -\ell(\ell+2) +L^2 E^2
 \ee
 Indeed the radial integral equation in this case can be solved exactly by taking
\bea
 {{R}}_{{\rm in}, \ell}(r)  &=& A_\ell \,  { J_{\ell} (E r) \over r^\alpha}  \nonumber\\
{{R}}_{\ell}(r)  &=& { A_\ell \over  2 r^\alpha }   \left(   H^1_{\nu_\ell} (E r)   + e^{2{\rm i} \delta_\ell} \,      H^2_{\nu_\ell} (E r)     \right) 
\eea
 with $A_\ell$ arbitrary constants specifying the incoming wave 
 and
 \bea
\delta(\ell,E) &=& {\pi \over 2} (\ell+1- \sqrt{(\ell+1)^2-E^2 L^2 }   )\nonumber\\
 \nu_\ell &=&\sqrt{(\ell+1)^2-E^2 L^2 }  
\eea
 The phase shift can then be written as
 \be
 e^{2{\rm i} \delta(\ell,E) }= e^{ {\rm i}\, \pi  \, (\ell-\ell_L) }\approx   e^{  {\rm i}\, \pi  (\ell- \sqrt{\ell^2-L^2\, E^2} ) }
 \ee
  Identifying $\ell=b E$ and using (\ref{thetadelta}) one finds the deflection angle
   \be
\theta=- {2\over E } {\partial \delta(E,b)  \over \partial b}  = \pi \left[    { b\,  \over  \sqrt{b^2 -L^2}   }-1 \right]  \label{thetadeltaf}
\ee
  in perfect agreement with the result  (\ref{geod5}).  
  
   \subsection{D1D5 fuzzball case}
   
 The Klein-Gordon equation in the fuzzball geometry reads 
 \begin{align}
& \square \Phi({\bf x}, z,t )=  \left[ {  \partial_\r \left( \r (\r^2+a^2 ) \partial_\r \right)  \over H  \r(\r^2+a^2 \cos\vartheta^2 ) }+
\frac{  \partial_\vartheta \left( \sin(2\vartheta)
  \partial_\vartheta \right)  }{ H (\r^2+a^2\cos\vartheta^2)\, \sin(2\vartheta)}  \right.  \label{kgfield} \\
 &\left. +H (\partial_z^2-\partial_t^2)  +
 \frac{(\partial_\vf+\partial_t \w_\vf)^2}{H (\rho^2+a^2)\sin^2_\vartheta  }+    \frac{(\partial_\y- \partial_z \w_\y)^2}{H  \r^2 \, \cos^2_\vartheta  }  \rg] \Phi({\bf x}, z,t ) =0\nonumber
 \end{align}
  The scalar wave $\Phi({\bf x}, z,t )$ can be conveniently expanded in 3-sphere  harmonics (see Appendix for details)
  \be
 \Phi({\bf x}, z,t)=  e^{-{\rm i} E t +{\rm i} P\, z}  \sum_{\ell=0}^\infty \sum_{m'=-{\ell\over 2} }^{\ell\over 2}\sum_{m=-{\ell\over 2} }^{\ell\over 2} D^{\ell\over 2}_{  m' m  } (\Omega)
 \,R^{\ell}_{m' m} (\rho)
 \ee
 with  $\Omega=(\varphi,\psi,\vartheta)$ the coordinates on the sphere, $D^{\ell/2}_{  m' m  } (\Omega)$ the eigenvectors of the flat Laplacian on the 3-sphere
   \bea
 \nabla^2_{S^3} D^{\ell\over 2}_{m'm} (\Omega) &=& \left[ \frac{  \partial_\vartheta \left( \sin(2\vartheta)
  \partial_\vartheta \right)}{ \sin(2\vartheta)}   -  \frac{ (m-m')^2}{   \sin^2_\vartheta  }-\frac{(m'+m) ^2}{  \cos^2_\vartheta } \right]  D^{\ell\over 2}_{m' m }(\Omega) \nonumber\\
&=&  -\ell(\ell+2)\, D^{\ell\over 2}_{m' m }(\Omega)  \eea
$D^{\ell\over 2}_{  m' m  } (\Omega)$ are the matrix elements of finite $SO(3)\sim SU(2)$ rotations in the $j=\ell/2$ representation and can be written as
 \be
 D^{\ell\over 2}_{m'm}  (\Omega) = e^{{\rm i}  (m-m') \varphi +{\rm i} (m+m')\,  \psi}      
 d^{\ell\over 2}_{ m' m }(2\vartheta)
 \ee
 in terms of the Wigner d-matrix  (see Appendix for details) after the identification
  \be
  J_\psi={m+m' }  \qquad , \qquad     J_\varphi={m-m'} 
  \ee  
The Klein-Gordon equation then becomes
   {\small
 \begin{align}
&  \left[ {  \partial_\r \left( \r (\r^2+a^2 ) \partial_\r \right)  \over   \r(\r^2+a^2 \cos\vartheta^2 ) }+
\frac{  \partial_\vartheta \left( \sin(2\vartheta)
  \partial_\vartheta \right)  }{  (\r^2+a^2\cos\vartheta^2)\, \sin(2\vartheta)} +H^2 (E^2-P^2) \right.  \label{kgfield} \\
 &\left.   -
 \frac{(m{-}m'{+}E \w_\vf)^2}{ (\rho^2+a^2)\sin^2_\vartheta  }-    \frac{(m{+}m'{-} P \w_\y)^2}{  \r^2 \, \cos^2_\vartheta  }  \rg] D^{\ell\over 2}_{  m' m  } (\Omega)
 \,R^{\ell}_{m' m} (\rho)
 =0\nonumber
 \end{align}
 }

 We notice that thanks to the $U(1)_\psi \times U(1)_\varphi$ isometry of the fuzzball metric, the Laplacian operator in
  (\ref{kgfield}) does not mix components with different $m$ (or $m'$), so the equations for the modes labeled by $(m,m')$ can be solved separately. This is clearly not the case for the 
  $\ell$-modes that mix with one another in the scattering process. As a result, in order to determine the phase shift, one should in principle diagonalize the wave operator and identify the exact eigen-modes. This is the wavy analogue of the problem that one faces solving the geodetic equations in the fuzzball geometry. As before, notwithstanding the separability of the problem, for simplicity, we will focus on the cases 
\begin{itemize}

\item{  $m=m'>>1$, {\it i.e.} $J_\psi = 2m$, $J_\varphi = 0$.  The $\ell$-sum is dominated by the mode $\ell=2m$. The Wigner d-matrices reduce to  
  \be
  d^{m}_{ m,m }(2\vartheta) = (\cos\vartheta)^{2m}
  \ee
  and are peaked at $\vartheta=0$. 
  }

\item{  $m=-m'>>1$, {\it i.e.} $J_\varphi = 2m$, $J_\psi = 0$.  The $\ell$-sum is dominated by the mode $\ell=2m$. The Wigner d-matrices reduce to  
  \be
  d^{m}_{{-}m,m }(2\vartheta) = (\sin\vartheta)^{2m}
  \ee
  and are peaked at $\vartheta={\pi\over 2}$. 
  }

\end{itemize}

 In the following we consider the wave scattering in these two 2-planes.

 \subsubsection{Wave scattering for $J_\varphi=0$} 
  
For  $ m=m'>>1$ spherical modes the solution of the field equation is  peaked around $\vartheta=0$, whereby $\cos\vartheta \approx 1$. The wave equation becomes
 \begin{align}
0
 & =\left[  {1\over  \r^3 } \partial_\r \left( \r^3   \partial_\r \right) +{ 1\over \rho^2} \nabla^2_{S^3}   +E_P^2  -   \widehat{U}_{\rm eff} (\rho)
    \right]  R_{m,m}^{2m} (\rho)
 \end{align}
 with $E_P^2=E^2-P^2$ and 
 {\small
 \bea
\widehat{U}_{\rm eff} (\rho) &=&  -  {a^2 \over  \r^3 } \partial_\r (\r  \partial_\r )
 - E_P^2   \left[ H^2  \left(1{+}{a^2\over \rho^2} \right){-}1 \right] + {4 m^2\over \rho^2}  
  \left[    \left(1{-}  {P a L_1 L_5   \over 2m (\rho^2+a^2)} \right)^2    \left(1{+}{a^2\over \rho^2} \right){-}1 \right]       
    \nonumber \\
 &  \approx  & {-}  {a^2 \over  \r^3 } \partial_\r (\r  \partial_\r ) {-} {E_P^2( L_1^2{+}L_5^2{+}a^2 ) \over \rho^2}  - {4m P a L_1 L_5  -4 m^2 a^2\over \rho^4}  
        +\ldots     
 \eea
 }
 where in the last line we display the first few terms in the expansion of the potential at large $\rho$. Writing 
 \be
 \widehat{U}_{\rm eff} (\rho) =-\sum_{n,p}  {c_{n,p}\over \r^n}\partial_\r^p
  \label{ueffnp} 
  \ee
 one finds for the phase shift
 \be
 2\, \delta=\pi \sum_{n,p}  E_P^{n+p-2} \,  c_{n,p}\, I_{n,p}   \label{deltanp}
 \ee
 with
  \be
  I_{n,p}=\int_0^\infty    J_{\ell} (  y)   \partial^p_{y}  \left[   { J_{\ell} (  y )  \over  y } \right]   y^{2-n} dy
 \ee
 The  results of the integrals for the  relevant $(n,p)$  are displayed in the following table
 \bea
\begin{array}{c|cccc}
  (n,p) &   (2,2) & (3,1) &  (2,0) & (4,0)    \\
 \hline
   I_{np} & - {1\over 4 \ell}&  { (1-\alpha) \over 4 \ell^3}  & {1\over 2 \ell}  &  {1\over 4 \ell^3 }  \\  
\end{array}
 \eea
 Combining (\ref{ueffnp}) and (\ref{deltanp}), and taking $\ell=2m=b_\psi E$, $\alpha=1$ one finds
\be
2 \delta  \approx  {\pi E^2_P \over 2 E } \left(  {L_1^2{+}L_5^2   \over  b_\psi }  + {   P a L_1 L_5\over  b_\psi^2 \,  E  }     +\ldots      \right)        
\ee
leading to $\Delta\psi  \approx  - {2\over E } {\partial \delta   \over \partial b_\psi}$, i.e.
 \be
 \boxed{
\Delta\psi_{\rm fuzz}   \approx      {\pi  (1-v^2) \over 2 b_\psi^2 }     \left(      L^2_1+L^2_5    +{  2 a   L_1 L_5 \,   v \over  b_\psi}   +\ldots   \right)     
   \label{thetadeltacn2}
   }
\ee
in perfect agreement with what found using classical geodetic motion in the circular fuzzball geometry in the plane $\vartheta=0$.

 \subsubsection{Wave scattering for $ J_\psi=0$} 
  
The study  for $ m=-m' $ modes proceeds {\it mutatis mutandis} along our analysis above. In this case the spherical harmonic functions are peaked at $\vartheta={\pi \over 2}$. 
 The  effective potential then becomes
 {\small
 \bea
\widehat{U}_{\rm eff} (\rho) &=&  -  {a^2 \over  \r^3 } \partial_\r (\r  \partial_\r )
 - E_P^2   (H^2 {-}1 )  { -} {4 m^2\over \rho^2}  
  \left[    \left(1{+}  {E a L_1 L_5  \over 2m \rho^2} \right)^2    {\rho^2 \over \rho^2 + a^2} {-}1 \right]       
    \nonumber \\
 &  \approx  & -  {a^2 \over  \r^3 } \partial_\r (\r  \partial_\r )    {-} {E_P^2 ( L_1^2+L^2_5) \over \rho^2} { -} {4m E a L_1 L_5 -4 m^2 a^2  \over \rho^4}  
+\ldots     
 \eea
 }
Using again (\ref{deltanp})  now with  $\ell=2m=b_\varphi E$ one finds
 \be
2 \delta  \approx  {\pi E_P^2 \over 2 E } \left(  {L_1^2{+}L_5^2   \over  b_\varphi}  - {    a L_1 L_5\over  b_\varphi^2  }        \right)      +\ldots     
\ee
leading to $\Delta\varphi  \approx  - {2\over E } {\partial \delta   \over \partial b_\varphi}$, i.e. 
 \be
 \boxed{
\Delta\varphi_{\rm fuzz}  \approx     {\pi  (1-v^2)\over 2  b_\varphi^2}    \left(    L^2_1+L^2_5     -{  2 a   L_1 L_5 \,   \over      b_\varphi}      \right)       +\ldots 
   \label{thetadeltacn3}
   }
\ee
 in perfect agreement with the deflection angle found using classical geodetic motion in the circular fuzzball geometry in the plane $\vartheta=\pi/2$.

\section{ String amplitudes }
\label{stringscatt} 
 
  The scattering of massless closed-string probes in a D-brane background is described by a two-point amplitude on a string world-sheet with boundaries mapped onto the D-brane stack. In the high energy Regge limit, the amplitude is dominated by  diagrams involving the exchange of nearly on-shell gravitons  between the high-energy string state and the brane,  whose elementary block is the tree-level (disk) string-brane scattering amplitude {\it i.e.} the closed string two-point amplitude on the disk. In the leading eikonal approximation, the dominant diagrams resum to the exponential of the tree-level diagram
  \be
  S = e^{2{\rm i} \delta (s,b)}=e^{ {\rm i} \widehat{\cal A}_{\rm disk} (s,b) \over 2 E } + \dots
  \ee
 with $\widehat{\cal A}_{\rm disk} (s,b)$ the Fourier transform of the string amplitude ${\cal A}_{\rm disk} (s,t)$ {}\footnote{This is related to the wave scattering amplitude (\ref{wavescattamp}) by $f=E^{\alpha - {1\over 2}}{\cal A}$.} \be
2\delta (s,b) ={ \widehat{\cal A}_{\rm disk} (s,b) \over 2 E} = {1\over \sqrt{s}}  \int \frac{d^{d{-}1}{\bf q} }{(2\pi)^{d{-}1}}\, {\cal A}_{\rm disk}(s,-{\bf q}^2)  \, e^{ {\rm i} \, {\bf q}\cdot {\bf b}}
\ee
 with the $q$-integral running over the space inside ${\bf R}^d$ transverse to  vector ${\bf p}_1-{\bf  p}_2$. 
The functional relation between the deflection angle $\Delta\theta$ and the impact parameter $b$  was given in (\ref{thetadelta}), so one finds
\be
\boxed{
\Delta\theta\approx -  {2 \over E }\,{\partial \delta\over \partial b}\approx -  {1 \over  s }\,{\partial \widehat{\cal A}_{\rm disk}(s,b) \over \partial b} 
}  \label{derA}
\ee 
  In the remaining of this section we review the computation of the deflection angle in high energy scattering of a scalar particle from a Dp-brane  using both geodetic motion in the Dp-background and the string scattering amplitude. In the next section the results will be extended to the case of the D1-D5 fuzzball geometry.

 \subsection{Dp-branes}

  At the disk level, the scattering amplitude of a massless scalar particle   in the  NSNS sector is described by the two point amplitude
\be
 \left\langle     \, W(p_1,\tilde p_1) \, \, W(p_2,\tilde p_2) \right\rangle_{\rm disk} = {\cal A}_{\rm disk} (s,t) \, {\rm Tr} ( {\cal E} _1 \,  {{R}}_{Dp} \, {\cal E} _2 \, {{R}}_{Dp})  +\ldots  \label{amplitude}
\ee
with $W(p_i,\tilde p_i)$ the NS{-}NS closed string vertices with momenta 
 \be
 p_i=(E,\vec{k}_i,\vec{0}) \quad\quad\quad      \tilde p_i={{R}}_{Dp} \, p_i =(E,-\vec{k}_i,\vec{0})   \qquad   \vec{k}^2_1=\vec{k}_2^2=E^2 \quad ,
 \ee
${\cal E}_i$ the incoming and outgoing polarisation    
 tensors and
\be
{{R}}_{Dp}={\rm diag}(+1, -1_{9-p},+1_{p} )
\ee
 the boundary reflection matrix.  Notice that $\vec{k}_i$ are aligned along the Dirichlet directions of the Dp-brane target. We have isolated in (\ref{amplitude}) the kinematical structure  dominating elastic scattering in the high energy Regge limit and denoted by dots the rest. 
 The string amplitude ${\cal A} (s,t)$ is a function of the Mandelstam variables 
 \bea
 s &=& -2 \, p_1\, \tilde p_1= -2\, p_2\, \tilde p_2  =4\, E^2\nonumber\\
   t & =&  -2 \, p_1\,  p_2= -2\, \tilde p_1\, \tilde p_2 =- 4\, E^2\, \sin^2{\theta\over 2} 
  \eea
 with $\theta$ the scattering angle. 
Setting $\alpha'=2$, one finds\footnote{Here we use the Stirling formula $\Gamma(x)\sim e^{x(\log x-1)}$.}
 \bea
{\cal A}_{\rm disk} (s,t)  &  {{=}} & s\,   { \Gamma \left( -\frac{s}{2} \right)  \,\Gamma \left( -\frac{t}{2} \right)  \over \Gamma \left(-\frac{s}{2}-\frac{t}{2} \right)  }
 \underset{ s >>1 }{\approx }  \,\Gamma \left( -\frac{t}{2} \right) \, e^{{\rm i} \,\pi\, t\over 2} \, s^{1+{t\over 2}}   \label{ast}
 \eea
 where in the right hand side we kept only the leading term in the high energy limit $s>>1$. We are interested in the limit of large impact parameter $b>> \sqrt{\ln s}$ of the 
function  $\widehat{\cal A}(s,b)$ defined as the Fourier  transform  of (\ref{ast}). In this limit the integral is dominated by a saddle point at ${\bf q}\approx 0$ leading to\footnote{
In this limit the integral is almost real since its imaginary part  is exponentially suppressed when  $b>> \sqrt{\ln s}$. When $b$ is of order $\sqrt{\ln s}$, the imaginary part of the integral is dominated by a saddle point at  ${\bf q}_*={ {\rm i} \, {\bf b} \over \ln s}$ resulting into
\bea
{\rm Im}\, \widehat{\cal A}(s,b) 
   &  \sim &   {\rm i}  s\, \int  d^{d}{\bf q} \,  e^{ -{q^2\over 2}\ln s +{\rm i} {\bf q} \cdot {\bf b}  }    \sim  s\,  (\ln s)^{{-}{d\over 2}}  \, e^{-{b^2\over 2 \ln s}}  
 \eea
}
\bea
 \widehat{\cal A}_{\rm disk}(s,b) 
   =   \int {d^{8{-}p} {\bf q} \over (2\pi)^{8{-}p}} \, {\cal A}_{\rm disk} (s,-q^2)\,  e^{ {\rm i} {\bf q} \cdot {\bf b}}  \approx  
 s\,   \int  {d^{8{-}p} {\bf q} \over (2\pi)^{8{-}p}} \,  { e^{ {\rm i} {\bf q} \cdot {\bf b}} \over q^2}  
      \approx  { s \over  |b|^{6-p} }  \label{astringdp}
 \eea
 with  the integral running over the space traverse to the stack of Dp-branes and to the longitudinal momentum ${\bf p}_1{-}{\bf p}_2$.  Plugging this into (\ref{derA}) one finds that the deflection angle falls as $b^{p-7}$ at large $b$ as predicted by the geodetic motion analysis. 
 
  Specifying to a system of D1 and D5 branes on $\mathbb{R}^{1,4}\times T^5$, the integral in (\ref{astringdp}) runs over a 3-dimensional space, so one finds  
    \be
 \Delta\Theta_{D1}+  \Delta\Theta_{D5}  \sim       { \left(    L^2_1+L^2_5        \right)  \over    b^2}
 \ee 
reproducing the leading deflections of the geodesics in the fuzzball background for $\Theta=\psi,\varphi$.

\subsection{D1D5 fuzzball}

Let us consider the scattering of a NS-NS massless scalar from a D1D5 fuzzball. The binding of the two stacks of brane is produced by a condensate of massless open strings at zero momentum. At lowest order in perturbation theory the process is captured by a disk with mixed boundary conditions. The relevant amplitude reads 
\be
\left\langle    V_{\mu}( 0 ) \, V_{\bar \mu}( 0 )   \,   \, W(p_1,\tilde p_1) \, W(p_2,\tilde p_2) \right\rangle ={\cal A}^{D1D5}_{\rm disk} (s,t) 
    \, {\rm Tr} ( {\cal E} _1 \,  {{R}}_{{D1D5}} \, {\cal E} _2 \, {{R}}_{{D1D5}})  +\ldots  
\ee
and requires the insertion of two twisted open strings connecting D1- and D5-brane boundaries at zero momenta, described by the massless fermion vertex operators 
\bea
V_{\mu} (0)  = c\, \mu_A \,  e^{-\varphi/2}  \, S^A \, \sigma (z_1) \qquad , \qquad
V_{\bar \mu} (0)  = \int dz_2\, \bar \mu_B \,  e^{-\varphi/2}  \, S^B \, \sigma (z_2)   
\eea
where $\sigma$ denote the ${\bf Z}_2$ twist-field along the ND 4-plane, 
$S^{A/B}$ the $SO(1,5)$ spin fields, $\varphi$ the super-ghost boson and $c$ the bosonic ghost, and two NS{-}NS massless closed strings described by the vertex operators 
\bea
&&W(p_1,\tilde p_1) = \int dz_3 dz_4 \left( {{R}}_{{D1D5}}\cdot {\cal E}_1 \right)_{PQ} e^{- \varphi}  \psi^{Q} e^{i p_1 X}(z_3) 
 (\partial X^P - i \tilde{p}_1\psi \psi^P)  e^{i \tilde{p}_1 X} (z_4) \nonumber\\ 
&&W(p_2,\tilde p_2) = \left( {{R}}_{{D1D5}}\cdot {\cal E}_2 \right)_{MN} c (\partial X^M{-}i p_2\psi \psi^M)  e^{i p_2 X}(z_5)  c (\partial {X}^N{-}i \tilde{p}_2\psi \psi^N) e^{ i\tilde{p}_2 {X}} (z_6) \nonumber \\
\eea
where $X$ and $\psi$ denote the bosonic coordinates and their fermionic super-partners on the world-sheet.
The index $A$ labels the spinor representation of the $SO(1,5)$ Lorentz group of the space where D1 and D5 branes are `parallel' (NN or DD). 
We take the same kinematics as before and focus on open-string condensates in the ${\bf 10}$ of $SO(1,5)$ 
  \be
  {\cal O}^{PQR}=  \langle \bar\mu \Gamma^{PQR} \mu \rangle   
  \ee 
   There is no massless open string state in this representation, so such a condensate can only emit closed strings.   We focus on the following coupling
   \be
 \langle\bar\mu \Gamma^{MNP} \mu\rangle \,    p_{1M} \,p_{2N}  \tilde p_{2P}\,   {\rm Tr} ( {\cal E}_1 \,  {{R}}_{D1D5} \, {\cal E}_2 \, {{R}}_{D1D5})  =s\,  {\bf a}\cdot {\bf q} \,   {\rm Tr} ( {\cal E}_1 \,  {{R}}_{D1D5} \, {\cal E}_2 \, {{R}}_{D1D5}) 
   \ee
   with  ${\bf a}$ and $ {\bf q}$  defined by 
   \be
   {\bf a}^P= {1\over 2\, s} ( \mu \Gamma^{MNP} \mu) \,    (p_{1M}-p_{2M})  (p_{2N}+\tilde p_{2N})   \qquad , \qquad   {\bf q}={\bf p}_1+{\bf p}_2
   \ee
   We notice that  both ${\bf a}$ and $ {\bf q}$ are transverse to the longitudinal momentum ${\bf p}_1-{\bf p}_2$. 
The $1/s$ normalisation in the definition of ${\bf a}$ is included in order to ensures that ${\bf a}$ be finite in the large $s$ limit.  
 To evaluate the string amplitude we choose specific components for the spinors and polarisation tensors, let us say
 \be
 A=B=\left({1\over 2},{1\over 2},{1\over 2}\right)   \qquad , \qquad    (M,N)=(5,3)   \qquad , \qquad    (P,Q)=(\bar 5,\bar 3)
 \ee
where the (complex) directions $1,2,3$ ($\bar{1},\bar{2},\bar{3}$) are along $SO(1,5)\supset SO(1,1)\times SO(4)$ (NN,DD), while 
$4,5$ ($\bar{4},\bar{5}$) are along $SO(4)$ (ND).
For this choice, one can easily see that the terms $\partial X$'s do not contribute, thus the correlator to evaluate is
\begin{align}
\langle &(c \, e^{-\varphi/2}  \, e^{\frac{i}{2} \varphi_1+\frac{i}{2} \varphi_2+\frac{i}{2} \varphi_3} \, \sigma) (z_1)
(e^{-\varphi/2}  \, e^{\frac{i}{2} \varphi_1+\frac{i}{2} \varphi_2+\frac{i}{2} \varphi_3} \, \sigma) (z_2)
(e^{- \varphi}  \psi^5)(z_3) (\tilde{p}_1^1 \bar{\psi}^1 \psi^3)  (z_4) \, \times \nonumber \\
&( c \,  p_2^3 \bar{\psi}^3 \bar{\psi}^5)(z_5)  ( c\, \tilde{p}_2^2 \bar{\psi}^2 \bar{\psi}^3)(z_6) 
\rangle \, \langle e^{i p_1 X(z_3)} e^{i \tilde{p}_1 X(z_4)}  e^{i p_2 X(z_5)} e^{ i\tilde{p}_2 {X(z_6)}}  \rangle
\end{align}
The components $\tilde{p}_1^1$, $ p_2^3$ and $\tilde{p}_2^2$ have been chosen to avoid cuts intersecting the $x$-integration contour in the computations below. The amplitude takes the simple form
      \bea
{\cal A}^{D1D5}_{\rm disk} (s,t)  =  s\, {\bf a}\cdot {\bf q} \,    \int   { z^{1\over 2}_{15}  z^2_{56} \,  dz_2 \, dz_3\, dz_4         \over    (z_{25} z_{13}z_{23} )^{1\over 2}      z_{26} z_{35} z_{45} z_{4 6}  }     \, \left( {z_{34} z_{56}\over z_{36}z_{45}}  \right)^{s\over 2} 
\left({ z_{35} z_{46} \over z_{36}z_{45}}   \right)^{t\over 2}      
 \eea
Setting 
\be
z_1=0 \quad , \quad z_2=x \quad , \quad z_3=z \quad , \quad z_4=\bar z  \quad , \quad  z_5=i \quad , \quad  z_6=-i
\ee
  one finds
\bea
{\cal A}^{D1D5}_{\rm disk} (s,t)  =  s\, {\bf a}\cdot {\bf q} \,   \int    { dx \, d^2 z    (z-\bar z)^{-{s\over 2} } \,|z-i|^{-t}  \,  |z+i|^{s+t}  \over    (x+i) (\bar z-i) (\bar z+i) (z-i)   \sqrt{  (x-i)  \,   z \,  (x- z)  }   }     
\eea
 The integral over $x$ can be computed closing the contour along the lower half plane and picking the residue at $x=-i$, 
\bea
{\cal A}^{D1D5}_{\rm disk} (s,t)   =  s\, {\bf a}\cdot {\bf q} \,   \int       d^2 z    (z-\bar z)^{-{s\over 2} } \,  |z-i|^{-t-2}    \, |z+i|^{s+t-2 }   \, z^{-{1\over 2}}  \,  (z+i)^{{1\over 2}}   \nonumber
\eea
 In terms of the variable
  \be
w = {z-i \over z+i}
\ee
the final integral takes the form
 \bea
&& {\cal A}^{D1D5}_{\rm disk} (s,t) 
= s\, {\bf a}\cdot {\bf q} \,    \int_{\cal D}     d^2w   \left(  1-|w|^2   \right)^{-{s\over 2} }\,   |w|^{-t-2}\, (1+ w)^{ -{1\over 2} } \nonumber\\
&&=   2\pi\,   s\, {\bf a}\cdot {\bf q} \,     \int_0^1 dr   \left(  1-r^2   \right)^{-{s\over 2} }\,   r^{-t-1} =  \pi \,  s\, {\bf a}\cdot {\bf q}  
 \,  { \Gamma \left( 1-\frac{s}{2} \right)  \,\Gamma \left( -\frac{t}{2} \right)  \over \Gamma \left(1-\frac{s}{2}-\frac{t}{2} \right)  } 
 \eea
 with  ${\cal D}$ a unit disk centered at the origin.  We notice that in the large $s$ limit, this amplitude is exactly the same as the one describing the scattering from a Dp-brane except for the extra ${\bf a}\cdot {\bf q} $. Following the same steps as for the case of a Dp-brane stack one finds for the Fourier transform
 \bea
 {\cal A}^{D1D5}_{\rm disk}(s,b) 
  \approx 
 s\,   \int  {d^{3} {\bf q} \over (2\pi)^3} \,  {  {\bf a}\cdot {\bf q} \,    e^{ {\rm i} {\bf q} \cdot {\bf b}} \over q^2}  
      \approx  s\,    {  {\bf a}\cdot {\bf b} \over  |b|^{3} } 
 \eea
 leading to a deflection angle
 \be
 \boxed{
 \Delta \Theta_{D1D5}={  {\bf a}\cdot {\bf b} \over  |b|^{4} } 
 }
 \ee
 in perfect agreement, including the coefficient, with the geodetic results for $\Theta=\varphi,\psi$ after the identifications
 \be
 {\bf a} =
\left\{
\begin{array}{ccc}
  a L_1 L_5 v \, {\bf e}_\psi & {\rm for} & \vartheta=0  \\
 a L_1 L_5 \, {\bf e}_\varphi & {\rm for} & \vartheta={\pi\over 2}  \\
\end{array}
\right.
 \ee

\section{Conclusions}
\label{conclusions}

We have taken a first step towards the study of the dynamical properties of fuzz-balls. We have considered the simplest case of a D1D5 fuzzball with a circular profile function. 

We have studied geodetic motion, massless wave equation and string scattering in this background and checked consistency of the three approaches in the high energy (Regge) regime at large impact parameter. We consider this further success of the fuzzball proposal, including the peculiar behaviour at the critical impact parameter for capture of (massless) probes. 
 
It would be very interesting -- but quite challenging -- to extend the present analysis to non circular fuzzballs of two-charge systems if not even of  3-charge systems in five dimensions and of 4-charge systems in four dimensions. One can also consider scattering of massive, charged or (higher) spin probes in order to gain further insights in the dynamical properties of the fuzzball under consideration. 

Last but not least, it would be desirable to understand the connection between `critical' impact parameter, (apparent) loss of unitarity and opening of inelastic channels, that would be associated to the typical `size' or better `cross-section' of the fuzzball, that should (approximately) match the horizon area of the putative black-hole.

\section*{Acknowledgements}
  We acknowledge fruitful discussions with Andrea Addazi, Pascal Anastasopoulos, Guillaume Bossard, Ramy Brustein, Paolo Di Vecchia, Francesco Fucito, Stefano Giusto, Alfredo Grillo, Elias Kiritsis, Antonino Marcian\`o, Lorenzo Pieri, Gabriele Rizzo, Rodolfo Russo, Raffaele Savelli, Gabriele Veneziano, and Natale Zinnato. 
Part of the work was carried on while M.~B. and D.~C. were visiting the University of Crete in Heraklion and while M.~B. was visiting Fudan University in Shanghai. M.~B. and D.~C. would like to thank them for the kind hospitality. We thank the MIUR-PRIN contract 2015MP2CX4002 {\it ``Non-perturbative aspects of gauge theories and strings''} for partial support. 
\appendix

\section{Spherical Harmonics}
\label{sect:spherical_harmonics}

In this Appendix we  give some details on the special functions entering in the spherical harmonic  expansions of the field in the different   coordinate systems used in the main text.

\subsection{  $S^{d-1}$-harmonics}

 The metric of the sphere $S^{d-1}$ can be written as
  \be
 ds^2_{S^{d-1} } = d\theta^2+\sin^2\theta\, d\Omega_{d-2}
 \ee
With this choice the Laplace operator reads
 \be
\nabla^2_{S^{d-1}}= {1\over \sin^{2\alpha} \theta} \partial_\theta \left( \sin^{2\alpha} \theta\,\partial_\theta \right) \,  + \nabla^2_{S^{d-2}}
\ee
  A function $ f(\theta)$ can be in general expanded in $\theta$-harmonics 
  \be
  f(\theta)=\sum_{\ell=0}^\infty f_\ell \, C^{d-2 \over 2}_\ell (\cos \theta)
  \ee
  with  $C^\alpha_\ell (x)$ the Gegenbauer polynomials defined via 
\be
{1\over (1-2\, r\, x+r^2)^{\alpha}}=\sum_{\ell=0}^\infty C_\ell^\alpha(x)\, r^\ell    \label{gegen}
\ee
The Gegenbauer polynomials are eigenvectors of the Laplacian on $S^{2\alpha+1}$
 \be
\nabla^2_{S^{2\alpha+1}}\,  C^{\alpha}_\ell (\cos \theta) =- \ell(\ell+2\alpha) \, C^{\alpha}_\ell (\cos \theta)
\ee 
and can be written in terms of hypergeometric functions with $a=-n$ 
\be
C_n^\alpha(x)={\Gamma(2\alpha+1) \over \Gamma(n+1) \Gamma(2\alpha-n+1) } \,  {}_2 F_1 \left(-n,n+2\alpha,\alpha+{1\over 2}; {1-x\over 2} \right)
\ee
  In this basis the plane wave can be expanded as 
  \be
 e^{i E \, r\, \cos\theta} =   { 2^{\alpha } \, \Gamma(\alpha) \over (E r)^\alpha}  \sum_{\ell}^\infty  \,     {\rm i}^\ell (\ell+\alpha)     \,  J_{\ell+\alpha }(E\, r)      \, C^\alpha_\ell (\cos \theta) 
 \label{planewavexp}
\ee
  
\subsection{  $S^{3}$-harmonics}

 Being the group manifold of $SU(2)$, the 3-sphere $S^{3}$ is very special. Its metric can be written in the form
  \be
 ds^2_{S^3 } =       d\vartheta^2 {+}  \cos^2\vartheta\, d {\y}^2{+} \sin^2 \vartheta d {\vf}^2
\ee
In these variables the Laplacian operator  reads
 \be
 \nabla^2_{S^3} =  \frac{  \partial_\vartheta \left( \sin(2\vartheta)
  \partial_\vartheta \right)}{ \sin(2\vartheta)}   +  \frac{ \partial_\varphi^2}{   \sin^2_\vartheta  }+\frac{ \partial_\psi^2}{  \cos^2_\vartheta }  
  \ee
  Denoting by $\Omega=(\vartheta,\varphi,\psi)$ denoting the angular coordinates, a function $f(\Omega)$ on the 3-sphere can be expanded in the basis of harmonics 
  \be
 f(\Omega)=\sum_{\ell=0}^\infty \sum_{m'=-{\ell\over 2} }^{\ell\over 2}\sum_{m=-{\ell\over 2} }^{\ell\over 2} D^{\ell\over 2}_{  m' m  }(\Omega) f^\ell_{m' m}  
  \ee
  with   $D^{\ell\over 2}_{m' m }(\Omega)$  the eigenvectors of the 3-sphere Laplacian
  \be
 \nabla^2_{S^3}  D^{\ell\over 2}_{m' m }(\Omega) =  -\ell(\ell+2)\, D^{\ell\over 2}_{m' m }(\Omega)  
\ee
    The three sphere harmonics $D^{\ell\over 2}_{m' m }$ are given  by
  \be
 D^{\ell\over 2}_{m'm}  (\Omega) =  e^{{\rm i}  (m-m') \varphi +{\rm i} (m+m')\,  \psi}      
 d^{\ell\over 2}_{ m' m }(2\vartheta)
 \ee
 with
  \be
  d^j_{m' m}(2\vartheta) =\sqrt{ (j{+}m)!(j{-}m)! \over (j{+}m')!(j{-}m')! }\left(\cos{\vartheta}\right)^{m+m'}\left(\sin\vartheta \right)^{m-m'}
 P^{m-m',m+m'}_{\ell-m}(\cos 2\vartheta)
 \ee
the Wigner d-matrices and $P^{a,b}_{n}(x)$  the Jacobi polynomials given by
 \be
  P^{a,b}_{n}(x)={\Gamma(a+2) \over \Gamma(n+1) \Gamma(a+2-n) }  {}_2 F_1 \left(-n,n+a+b+1,a+1; {1-x\over 2} \right)
\ee 
The Wigner d-matrices are normalised according to
 \be
 \int  d\Omega\, D^{{\ell_1\over 2} *}_{ m'_1 m_1} (\Omega)\,  D^{{\ell_2\over 2}}_{ m_2' m_2}(\Omega)
 =  {4\pi^2 \over 2\ell+1} \,\delta_{\ell_1 \ell_2} \,\delta_{m'_1 m'_2} \,\delta_{m_1 m_2} 
 \ee
 The d-Wigner matrices are related to the Gegenbauer polynomials $C^1_\ell$ via the ``remarkable" addition formula
 \be
C_\ell^1( \Omega_{12} )= \sum_{m'=-{\ell\over 2} }^{\ell\over 2}\sum_{m=-{\ell\over 2} }^{\ell\over 2} D^{\ell\over 2}_{  m' m  }(\Omega_1)\, D^{\ell\over 2}_{  m' m  }(\Omega_2)
 \label{remarkable}
 \ee
with
\be
\Omega_{12}=\cos\vartheta_1 \, \cos\vartheta_2\,\cos( \psi_1-\psi_2) +  \sin\vartheta_1 \, \sin\vartheta_2\,\cos( \varphi_1-\varphi_2) 
\ee
the scalar product of two unit vectors in ${\bf R}^4$ pointing along $\Omega_1$ and $\Omega_2$ on the three sphere.  
We notice that since $D^{\ell\over 2}_{  m' m  }(\Omega)$ represent a finite $SO(3)\sim SU(2)$ rotation matrix in the ${j=\ell/2}$ representation, the above remarkable addition formula implies that  the Gegenbauer polynomials $C^1_\ell(x)$ are nothing but the characters of $SU(2)$ {\it viz.}
\be
C^1_\ell (\cos{\vartheta}) = \chi^{SU(2)}_{j= {\ell \over 2}} (2  \vartheta) = \sum_{m=-{j\over 2}}^{j\over 2}  e^{2 {\rm i} m \vartheta} 
\ee
Indeed, setting $x=\cos(\alpha/2)$, they both satisfy the same recursion relation 
\be
2x C_n^1(x) = C^1_{n+1}(x) + C^1_{n-1}(x)        \qquad  \leftrightarrow \qquad  2 \cos{\alpha\over 2} \chi_j(\alpha) = \chi_{j+{1\over 2}}(\alpha) + \chi_{j-{1\over 2}}(\alpha)
\ee
Using (\ref{remarkable}) together with (\ref{planewavexp}) one can write the harmonic expansion of a plane wave in 4-dimensional space as
 \be
 e^{i E \, r\, \Omega_{12} } =   {  1  \over E r }  \sum_{\ell=0}^\infty  \,  \sum_{m'=-{\ell\over 2} }^{\ell\over 2}\sum_{m=-{\ell\over 2} }^{\ell\over 2}    {\rm i}^\ell (\ell+1)     \,  J_{\ell+1 }(E\, r)      \, D^{\ell\over 2}_{  m' m  }(\Omega_1)\, D^{\ell\over 2}_{  m' m  }(\Omega_2) 
 \label{planewavexp}
\ee

\section{Schwarzschild metric}

For completeness and comparison, in this Appendix, we would like to review light-like geodesic, scattering with large impact parameter, free falling and critical impact parameter for capture in the Schwarzschild metric.
\begin{gather}
ds^2=- A(r) \, dt^2+A^{-1}(r)\, dr^2+r^2 (d\q^2+\sin^2 \q \, d \vf^2)\\
 A(r)= 1-\frac{2M}{r}
\end{gather}
We have two invariants: the energy and the angular momentum
\begin{equation}
-E=- A(r) \dot{t} \qquad
J_\vf=r^2 \sin^2 \q\, \dot{\vf}
\end{equation}
The angular equation allows us to fix $\q$.
\begin{align}
\ddot{\q}&+\frac{2}{r}\, \dot{\q}\, \dot{r}- \frac{\cos \q }{ \sin^3 \q}\frac{J_\vf^2}{r^2} = 0
\end{align}
If we set $\dot{\q}=0$, in order to maintain such condition for all $\t$, thus $\ddot{\q}=0$, we have two possibilities: $\q=\p /2$ or $J_\vf=0$. In the second case $\q$ can be freely chosen thus we can set it to $\q=\p /2$.

The radial equation can be replaced by the norm of the four-velocity
\begin{equation}
-\frac{E^2}{A}+ \frac{\dot{r}^2}{A} + \frac{J_\vf^2}{r^2}=0
\end{equation}
In order to study the trajectory we replace $\dot{r}$ introducing $\vf$ instead the geodesic parameter $\t$:
\begin{equation}
\dot{r}= \dot{\vf} \frac{dr}{d \vf} = \frac{J_\vf}{r^2} \frac{dr}{d \vf}
\end{equation}
The equation for the trajectory becomes
\begin{equation}
\frac{d\vf}{dr}=-\lf[\frac{A \, r^4}{b^2}-r^2\rg]^{-1/2}
\qquad b=\frac{J_\vf}{E}
\end{equation}
where we have introduced the impact parameter $b$.

Now we study in more detail the relation between the turning point $r_*$ and the impact parameter $b$ for a massless particle. The turning point is defined by the equation
\begin{equation}
\frac{r^2_*}{b^2}-1+\frac{2M}{r_*}=0 \quad \Rightarrow \quad r^3_*-b^2\,r_*+2 M b^2=0
\end{equation}
The polynomial is positive when $r=0$ and $r \to \infty$, thus if there are positive solutions they must appear in pair. The equation has not positive solutions below a critical value of $b$ which is determine when the pair of solutions degenerate into one, this happens when the discriminant vanishes
\begin{equation}
\Delta=4 M^2 \, b^4 (b^2-27 M^2)=0 \quad \Rightarrow \quad b_\textup{crit}=3 \sqrt{3} M
\end{equation}
At $b=b_\textup{crit}$ the particle stabilizes its motion into a circular orbit at $r=3M$. Below $b_\textup{crit}$ there are no positive intersections thus there is no turning point and the particle falls into the black hole.
 
To compute the leading contribution to the deflection angle we can use the formula \ref{resdp}, in which $n=1$ and $f_n=-2M$, obtaining Einstein's formula
\begin{equation}
\boxed{\Delta \theta= {2 M\over b} + \dots}
\end{equation}

ILet us consider now a radial fall, so we set $J_\vf=0$ and $\dot{\q}=0$. The geodesic equation becomes  $\dot{r}=- E$. Until $r$ vanishes we can use it as parameter to describe the geodesic. The time felt by an observer at large distance can be computed integrating the equation in $\dot{t}$:
\begin{equation}
\frac{dt }{d\t}=- E\frac{dt }{d r}=\frac{E}{1-\frac{2M}{r}}
\end{equation}
Integrating the equation we obtain
\begin{equation}
t_F-t_I=- \lf[r_F-r_I+2M \log \frac{r_F-2M}{r_I-2M}\rg]
\end{equation}
As known for $r_F$ that approaches the horizon $2M$, the time becomes infinite.

\providecommand{\href}[2]{#2}\begingroup\raggedright\endgroup
\end{document}